\documentclass{andp2012}% no class options needed by now
\usepackage[english]{babel}
\usepackage[utf8]{inputenc}
\usepackage[T1]{fontenc}
\usepackage{hyperref}
\title{Second-second moments and two observers testing quantum nonlocality}
\author[A. Bednorz]{Adam Bednorz\footnote{E-mail:~\textsf{abednorz@fuw.edu.pl}}}
\address{Faculty of Physics, University of Warsaw, ul. Pasteura 5, PL02-093 Warsaw, Poland}
\shortauthors{A. Bednorz}

\begin{abstract}
  We show that rejection of local realism in quantum mechanics can be tested by Bell-type inequalities for two observers and low-order moments of continuous and unbounded observables.
  We prove that one requires three observables for each observer for a maximally entangled state and  two observables for a non-maximally entangled state and write down appropriate inequalities and show violation by quantum examples.
  Finding an example for quadratures or position and momentum is left as an open problem.

\end{abstract}

\shortabstract
\begin{document}
\maketitle 

\section{Introduction}
\label{sec:intro}

Local realism means that outcomes of measurements by remote observers exist separately for each observer before the measurement is chosen.
It has been initially discussed by Einstein, Podolsky and Rosen (EPR) \cite{epr} in the context of measuring  position and momentum of an entangled state.
However, later Bell \cite{bell}, Clauser, Horne, Shimony, and Holt (CHSH) \cite{chsh} found a simple violation of local realism in a simple entangled state of two spins while measuring spin along different axes, with dichotomic outcomes. Despite the simplicity of the Bell model, it took over 50 years to confirm violation  \cite{hensen, vien, nist, munich} although the assumptions of the experiments 
require further research \cite{ab17}. On the theoretical side, many examples how to reject local realism have been proposed \cite{rmp}, including many observers \cite{ghz} or outcomes \cite{cglmp}.
The outcome can be a real number from continuous range, a result of position/momentum measurement like in the EPR case \cite{leon, gil, yur, weg, aub}. 

In this paper, we focus on a special class of tests local realism, involving moments  $\langle A^kB^l\rangle$ for two separated observers $A$ and $B$, with a given maximal degree $k+l$ and unbounded continuous variables.  Note that commonly used dichotomy $A=\pm 1$ is equivalent to the fourth-moment constraint $\langle(A^2-1)^2\rangle=0$.
The moment-based tests have been proposed first by Cavalcanti et al. \cite{cfrd}, involving ten observers, later reduced to three observers \cite{qhe}.
The original CHSH inequality can be rewritten in terms of up to fourth moments \cite{bb11}. Rejection of local realism needs always at least 4th moments \cite{bbb} for unbounded variables.

Tests of local realism with moments of continuous variables are useful when strong, projective measurements are hard or infeasible. Then weak measurements are more appropriate but at the cost of large noise added to the statistics
\cite{aharonov88, bb10, curic}. The sharp, discrete clicks are replaced by slightly shifted Gaussian distributions. One can reveal the underlying quantum statistics by subtracting the dominating Gaussian detection noise, or adding more detectors,
which is more efficient for low order moments and correlations. This is the case of optical \cite{pryde} and condensed matter attempts to test local realism \cite{loss1,loss2,kawa,lesovik}, measuring the flow of charges in mesoscopic junctions.
The low-order moments in tests of local realism can be useful also in relativistic quantum field theories where sharp measurement cause problems with renormalization \cite{abrel}, while moments and correlations can be regularized to avoid infinities.

The aim of the paper is to show how local realism can be rejected in an experiment involving two observers and measuring moments of the type $\langle A^kB^l\rangle$ with $k,l\leq 2$, i.e. second-second order. It is known that a natural class of inequalities involving such moments is satisfied both in quantum and classical mechanics \cite{salles}. We explored a general class of inequalities 
constructing a positive polynomial being a sum of low order monomials of jointly measurable observables. The violation of the positivity of the average of the polynomial implies the rejection of local realism. We show that such polynomial is not necessarily a sum of squares. Surprisingly, a maximally entangled state
requires at least three observables for each observer. However, there exists a class of examples involving non-maximally entangled states and only two observables at each side. Unfortunately, we have not found an example involving only position and momentum (quadratures).

\section{Motivation -- weak measurements}

Unlike in quantum optics, where most detections are click-based, measurements in solid state devices, such as tunnel junctions, quantum point contacts, dots, with semiconductors, superconductors, or in quantum Hall regime, are current-based \cite{loss1,loss2,kawa,lesovik}. It means that the flow of electrons is measured not by click but amplifying tiny voltage measured across the probe. Strong, projective measurements are infeasible because they would be too disturbing for the system. The outcome is then not $0$ or $1$ but a continuous value of voltage/current. Its statistics is dominated by Gaussian distribution, due to large amplification. The quantum signal is a small shift of the distribution.
It can be described in terms of weak measurements, where the detector interacts with the system instantly but also weakly \cite{aharonov88, bb10, curic}.
The simplest model of weak measurements uses Gaussian positive operator valued measure with Kraus operators \cite{kraus}
\begin{equation}
\hat{K}(a)=(2g/\pi)^{1/4}\exp(-g(\hat{A}-a)^2)\,,\label{kaa}
\end{equation}
where $g$ is the strength of the measurement of the operator $\hat{A}$ with the outcome $a$.
In the limit $g\to 0$ we have $\hat{K}(\pi2g)^{1/4}\to \hat{1}$ so there is no measurement at all (no dependence on $a$).
The actually measured probability $p'(a)=\mathrm{Tr}\hat{K}(a)\hat{\rho}\hat{K}^\dag(a)$ of the outcome $a$ at the state $\hat{\rho}$ has a form of convolution
\begin{eqnarray}
&&p'(a)=\int D(a-A)p(A)dA,\\
&& p(a)=\langle \delta(A-\hat{A})\rangle=\mathrm{Tr}\delta(A-\hat{A})\hat{\rho},\nonumber
\end{eqnarray}
with the dominating detection noise $D(a)=\sqrt{2g/\pi}e^{-2ga^2}$, with $\langle a^2\rangle_D=1/4g$, diverging at $g\to 0$. Here $p(A)$ is the probability of the outcome $A$ in the case of a strong, projective measurement $g\to \infty$ ($p(A)=\lim_{g\to\infty} p'(A)$), to which the noise $D$ is added. The advantage of weak measurements is their low invasiveness, i.e. the state after the measurement reads
\begin{equation}
\int da\hat{K}(a)\hat{\rho}\hat{K}^\dag(a)=\exp(-g\check{A}^2/2)\hat{\rho},
\end{equation}
with $\check{A}\hat{X}=[\hat{A},\hat{X}]$, so only off-diagonal elements of $\hat{\rho}$ in the eigenbasis of $\hat{A}$ are decreased and this effect is proportional to $g$.

To retrieve $p$ from $p'$, one has to make deconvolution, which is a terrible task, requiring Fourier transform of $p'$ and back. A simpler approach involves only moments i.e.
\begin{equation}
\langle a\rangle_{p'}=\langle A\rangle_p,\:\langle a^2\rangle_{p'}=\langle A^2\rangle_p+1/4g.
\end{equation}
If a second separate observer makes measurement of $\hat{B}$ (compatible with $\hat{A}$, i.e. $\hat{A}\hat{B}=\hat{B}\hat{A}$) with the outcome $b$ then the joint Kraus operator reads 
\begin{equation}
\hat{K}(a,b)=(2g/\pi)^{1/2}\exp(-g(\hat{A}-a)^2-g(\hat{B}-b)^2)\,,\label{kab}
\end{equation}
with the outcome probability
\begin{eqnarray}
&&p'(a,b)=\int D(a-A)D(b-B)p(A,B)dAdB,\nonumber\\
&& p(A,B)=\langle \delta(A-\hat{A})\delta(B-\hat{B}),\rangle
\end{eqnarray}
where again $p$ corresponds to strong, projective results.
The correlations with respect to $p'$ and $p$ are related
\begin{eqnarray}
&&\langle ab\rangle_{p'}=\langle AB\rangle_p,\;\langle a^2b\rangle_{p'}=\langle A^2B\rangle_p+\langle B\rangle_p/4g,\nonumber\\
&&\langle a^2b^2\rangle_{p'}=\langle A^2B^2\rangle_p+\langle A^2\rangle_p/4g+\langle B^2\rangle_p/4g+1/16g^2.
\end{eqnarray}
It is clear from the above relations that higher moments/correlations will involve high powers of $1/g$, which is diverging in the weak limit $g\to 0$. This is why keeping the order of moments/correlation low is desired from practical point of view.

\begin{figure}[t]
  \centering
  \includegraphics[width=\columnwidth]{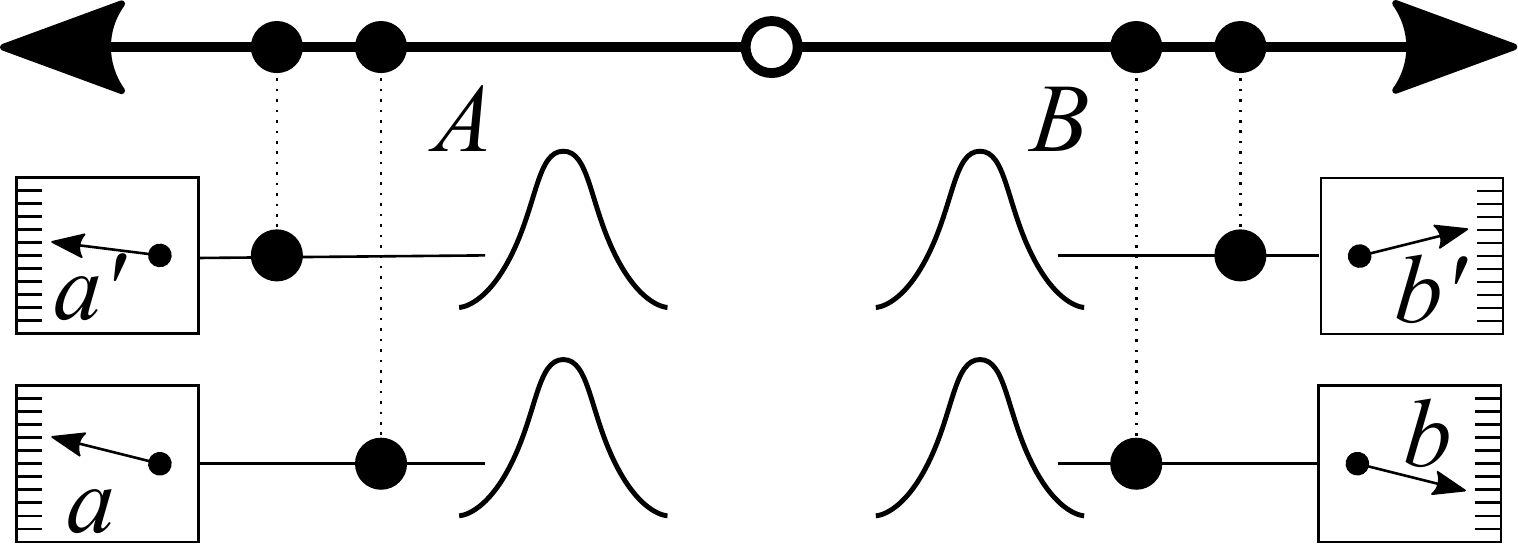}
  \caption{Weak measurement of up to second-second moments by maximally four detectors, coupled weakly to one of the parties, $A$ or $B$. Note that the outcome $a$ may differ from $a'$ (and $b$ from $b'$) because of independent random noise added to the intrinsic quantum value.
}
  \label{weakab}
\end{figure}

An alternative approach does not require subtraction of detection noise but measuring twice the same observable by two
identical and independent detectors. 
For a single party the Kraus operator reads
\begin{equation}
\hat{K}(a,a')=(2g/\pi)^{1/2}\exp(-g(\hat{A}-a)^2-g(\hat{A}-a')^2)\,.\label{kaap}
\end{equation}
Then the outcome probability reads
\begin{equation}
p'(a,a')=\int D(a-A)D(a'-A)p(A)dA.
\end{equation}
In this case 
\begin{equation}
\langle a\rangle_{p'}=\langle a'\rangle_{p'}=\langle A\rangle_p,\:\langle aa'\rangle_{p'}=\langle A^2\rangle_p.
\end{equation}
The correlation $\langle aa'\rangle$ does not contain the noise because the detectors are uncorrelated.
This idea generalizes to two parties using four detectors altogether
as depicted in Fig. \ref{weakab}. The full Kraus operator reads
\begin{eqnarray}
&&\hat{K}(a,a',b,b')=(2g/\pi)\exp(-g(\hat{A}-a)^2-g(\hat{A}-a')^2)\times\nonumber\\
&&\exp(-g(\hat{B}-b)^2-g(\hat{B}-b')^2)\,,\label{kabp}
\end{eqnarray}
with the outcome probability
\begin{eqnarray}
&&p'(a,a',b,b')=\\
&&\int D(a-A)D(a'-A)D(b-B)D(b'-B)p(A,B)dAdB.\nonumber
\end{eqnarray}
The correlations read
\begin{eqnarray}
&&\langle ab\rangle_{p'}=\langle a'b\rangle_{p'}=\langle ab'\rangle_{p'}=\langle a'b'\rangle_{p'}=\langle AB\rangle_p,\\
&&\langle aa'b\rangle_{p'}=\langle aa'b'\rangle_{p'}=\langle A^2B\rangle_p,\;
\langle aa'bb'\rangle_{p'}=\langle A^2B^2\rangle_p.\nonumber
\end{eqnarray}
Measurements of third moments of electric current current in the mesoscopic junction are very hard experimentally \cite{reulet1,reulet2,reznikov,reulet3} while measurements of fourth moments have  not yet succeeded.
Of course, subtracting the large noise or splitting $a$ into $a$ and $a'$ opens formally a loophole when testing local realism, but (i) the noise (also applied to $a-a'$) is well identified and there is no reason to take it into account to support local realism (ii) even  condensed matter tests of local realism subtracting this noise are still challenging \cite{palacios-laloy:10,white}. In the next sections all moments will be calculated with respect to $p$, assuming 
the deconvolution procedure $p'\to p$ has been already performed as described above.

\section{Moment-based inequalities and local realism}

The concept of local realism applies to two (at least) observers that make choices $x=0,1,2...$ and $y=0,1,2...$, respectively, for which they obtain probability $p(A,B|x,y)$ for the respective results $A$ and $B$. Since the observers and their choices are separate, the no-signaling principle says that
$p(A|x,y)\equiv p(A|x)$ cannot  depend on $y$ (and vice-versa for $B$). 
Local realism means that there exist hidden variables $A_x$, $B_y$ (locality means that there is neither $A_y$ nor $A_{xy}$) and a general positive probability $\tilde{p}(\{A_x\},\{B_y\})$  \cite{fine} such that
\begin{equation}
p(A,B|x,y)=\langle \delta(A-A_x)\delta(B-B_y)\rangle_{\tilde p}.
\end{equation}
In other words, the hidden variables $A_x$ and $B_y$ are revealed by the measurement at given choices.
Measurable correlations (moments) are related 
\begin{equation}
\langle A^kB^l\rangle_{xy}=\langle A_x^kB_y^l\rangle_{\tilde p}.
\end{equation}
We shall drop the index $\tilde p$ from now on for simplicity.

Contrary to the traditional Bell test, we do not impose any constraints on $A,B$ like dichotomy. They can be arbitrary real numbers. The concept of moment-based inequalities relies on construction of inequality
 involving measurable moments of $A,B$, i.e. $\langle A^k_xB^l_y\rangle$ with natural $k,l$, valid for arbitrary positive $\tilde{p}$. Measurability excludes correlations of
 different choices e.g. $\langle A^j_0A^k_1 B^m_y\rangle$ for $j,k\neq 0$. The first such inequality has been proposed by Cavalcanti et al. \cite{cfrd} reading
\begin{eqnarray}
&&\langle A_1^2B_1^2\rangle+\langle A_2^2B_1^2\rangle+\langle A_1^2B_2^2\rangle+\langle A_2^2B_2^2\rangle\geq\nonumber\\
&&
(\langle A_1B_1\rangle-\langle A_2B_2\rangle)^2+
(\langle A_1B_2\rangle+\langle A_2B_1\rangle)^2.\label{abe}
\end{eqnarray}
The quantum test of such inequality requires identification of moments with operator averages
\begin{equation}
\langle A^k_xB^l_y\rangle=\langle\hat{A}^k_x\hat{B}^l_y\rangle=\mathrm{Tr}\hat{\rho}\hat{A}^k_x\hat{B}^l_y,\label{qid}
\end{equation}
assuming Hermitian $\hat{A}_x$ and $\hat{B}_y$ acting in the tensor space $\mathcal H_A\otimes\mathcal H_B$ on its component, i.e.
$\hat{A}_x\to\hat{A}_x\otimes\hat{1}$ and $\hat{B}_y\to\hat{1}\otimes\hat{B}_y$, with the quantum state $\hat{\rho}$ represented by Hermitian, semipositive density matrix, normalized to 1. Here $\hat{A}_x=\hat{U}_x\hat{A}\hat{U}^\dag_x$ means measuring $\hat{A}$ given the local choice $x$ represented by unitary $\hat{U}_x\to\hat{U}_x\otimes\hat{1}$, and similarly $\hat{B}_y$. Unfortunately, (\ref{abe}) holds also in quantum mechanics, which is not trivial to prove \cite{bbb,salles}. Moreover, it is proved in \cite{salles} that every inequality
\begin{equation}
\sum_i\left\langle \sum_{xy}t_{ixy}A_xB_y\right\rangle^2\leq \sum_{xy}\langle A_x^2B_y^2\rangle,\label{cla}
\end{equation}
with constants $t_{ixy}$
holds in quantum mechanics under the identification (\ref{qid}) if it is true classically.
The class of such inequalities has (i) scaling property, $A_x,B_y\to \lambda A_x,\lambda B_y$ do not change it for an arbitrary real $\lambda$, (ii) equal coefficients at $\langle A_x^2B_y^2\rangle$ (independent of $x,y$), and (iii) no
terms $\langle A_x^2B_y\rangle$, $\langle A_xB_y^2\rangle$.

Nevertheless, already (\ref{abe}) generalized to three observers can be violated \cite{qhe}. Here we stick to two observers, $A$ and $B$.
One can rewrite standard CHSH inequality in terms of moments $\langle A_x^kB_y^l\rangle$ with $k+l\leq 4$. However, it involves pure fourth moments
$\langle A_x^4\rangle$ \cite{bb11}. The goal of this paper is to find an inequality involving only second-second order moments, namely $\langle A_x^kB_y^l\rangle$ with $k,l\leq 2$.
The gain is that only the observable and its square appear in the correlation, avoiding high order diverging terms, hard to eliminate in weak measurement approach or relativity.

We search of an appropriate inequality by examining positive polynomials, i.e.
\begin{equation}
W(\{A_x\},\{B_y\})\geq 0
\end{equation}
for all $A_x,B_y$ while the expansion of $W$ into monomials gives only terms $A_x^kB_y^l$ with $k,l\leq 2$.
In this way, such monomials do not contain products like $A_1A_2$, which cannot be  jointly measured.
Then the classical inequality $\langle W\rangle\geq 0$ holds for a nonnegative probability $\tilde{p}$ and can be tested in quantum mechanics. Note that $W$ is not necessarily a sum of squares of polynomials, for example
\begin{eqnarray}
&&A_1^2+A_2^2+B_1^2+B_2^2+(A_1^2+A_2^2)(B_1^2+B_2^2)\nonumber\\
&&
-\frac{3\sqrt{3}}{4}((A_1^2-A_2^2)(B_1+B_2)+(B_1^2-B_2^2)(A_1+A_2)).\label{nse}
\end{eqnarray}
The proof of positivity and impossibility of decomposition into polynomial squares is given in Appendix \ref{appa} (compare also with Choi example \cite{choi}).
Unfortunately, we have not found any quantum violation of (\ref{nse}), yet we failed to prove that the inequality holds in the general quantum cases.
Nevertheless, in the next sections, we show that the violating cases exist but the polynomials, inequalities, and violating states and observables are complicated.

\section{Maximally entangled state -- three choices}

First note that we can reduce the discussion to pure states i.e. $\hat{\rho}=|\psi\rangle\langle\psi|$. Otherwise 
\begin{equation}
\hat{\rho}=\sum_i q_i|\psi_i\rangle\langle \psi_i|,
\end{equation}
with $\langle \psi_i|\psi_j\rangle=\delta_{ij}$ and $q_i\geq 0$, $\sum_iq_i=1$
but also
\begin{equation}
\langle A^k_xB^l_y\rangle=\sum_i q_i\langle \psi_i|\hat{A}^k_x\hat{B}_y^l|\psi_i\rangle.
\end{equation}
If a positive $p_i$ exists for each pure state $|\psi_i\rangle$ and gives up to second-second moments  as predicted by quantum mechanics then $\sum_iq_ip_i$ will be the final probability.

Focusing on pure states, for two observers we can make Schmidt (singular value) decomposition
\begin{equation}
|\psi\rangle=\sum_j \phi_j|jj\rangle\label{schm}
\end{equation}
in certain tensor basis $|ij\rangle\equiv |i\rangle_A\otimes|j\rangle_B$ with real nonnegative $\phi_j$ satisfying $\sum_j\phi_j^2=1$.
For a maximally entangled state $\phi_j=1/\sqrt{N}$ for all $j$ where $N$ is the number of basis states in the decomposition. Note that the dimension of 
$\mathcal H_A$ and/or $\mathcal H_B$ can be larger than $N$, i.e. some basis states may not appear in the decomposition.
While maximally entangled states give the largest violation of CHSH or other inequalities, here counterintuitively they are useless if any of the observers, $A$ or $B$, has only two choices. In this case, one can explicitly construct the local probability $\tilde{p}$, see Appendix \ref{appb}.

We construct a minimal example for a violation requiring at least 3 choices for each observer.
The inequality, valid classically, reads
\begin{align}
&\langle A_1^2B_2^2\rangle+\langle A_2^2B_3^2\rangle+\langle A_3^2B_1^2\rangle+\nonumber\\
&2\sqrt{\langle A_1^2B_3^2\rangle\langle A_2^2B_2^2\rangle}
+2\sqrt{\langle A_2^2B_1^2\rangle\langle A_3^2B_3^2\rangle}
\nonumber\\
&+2\sqrt{\langle A_3^2B_2^2\rangle\langle A_1^2B_1^2\rangle}\geq 2(\langle A_1B_2\rangle+\langle A_2B_3\rangle+\langle A_3B_1\rangle)-1,\label{in33}
\end{align}
with all correlations  measurable. The inequality does not belong to the class of inequalities (\ref{cla}), because it does not satisfy its properties (i) and (ii).

To prove (\ref{in33}), note that
the following classical inequality holds
\begin{equation}
\langle(A_1B_2+A_2B_3+A_3B_1-1)^2\rangle\geq 0
\end{equation}
for all real numbers $A_x$, $B_y$. On the other hand opening squares we can reduce it to
\begin{align}
&\langle A_1^2B_2^2\rangle+\langle A_2^2B_3^2\rangle+\langle A_3^2B_1^2\rangle+\nonumber\\
&
2(\langle A_1B_3A_2B_2\rangle+\langle A_2B_1A_3B_3\rangle+\langle A_3B_2A_1B_1\rangle)
\\
&\geq 2(\langle A_1B_2\rangle+\langle A_2B_3\rangle+\langle A_3B_1\rangle)-1.\nonumber
\end{align}
Using Cauchy-Bunyakovsky-Schwarz (CBS) inequality we get
\begin{equation}
\sqrt{\langle A_2^2B_1^2\rangle\langle A_3^2B_3^2\rangle}\geq \langle A_2B_1A_3B_3\rangle
\end{equation}
and two others by cyclic shift of 123, and finally (\ref{in33}).

\begin{figure}[t]
  \centering
  \includegraphics[width=.5\columnwidth]{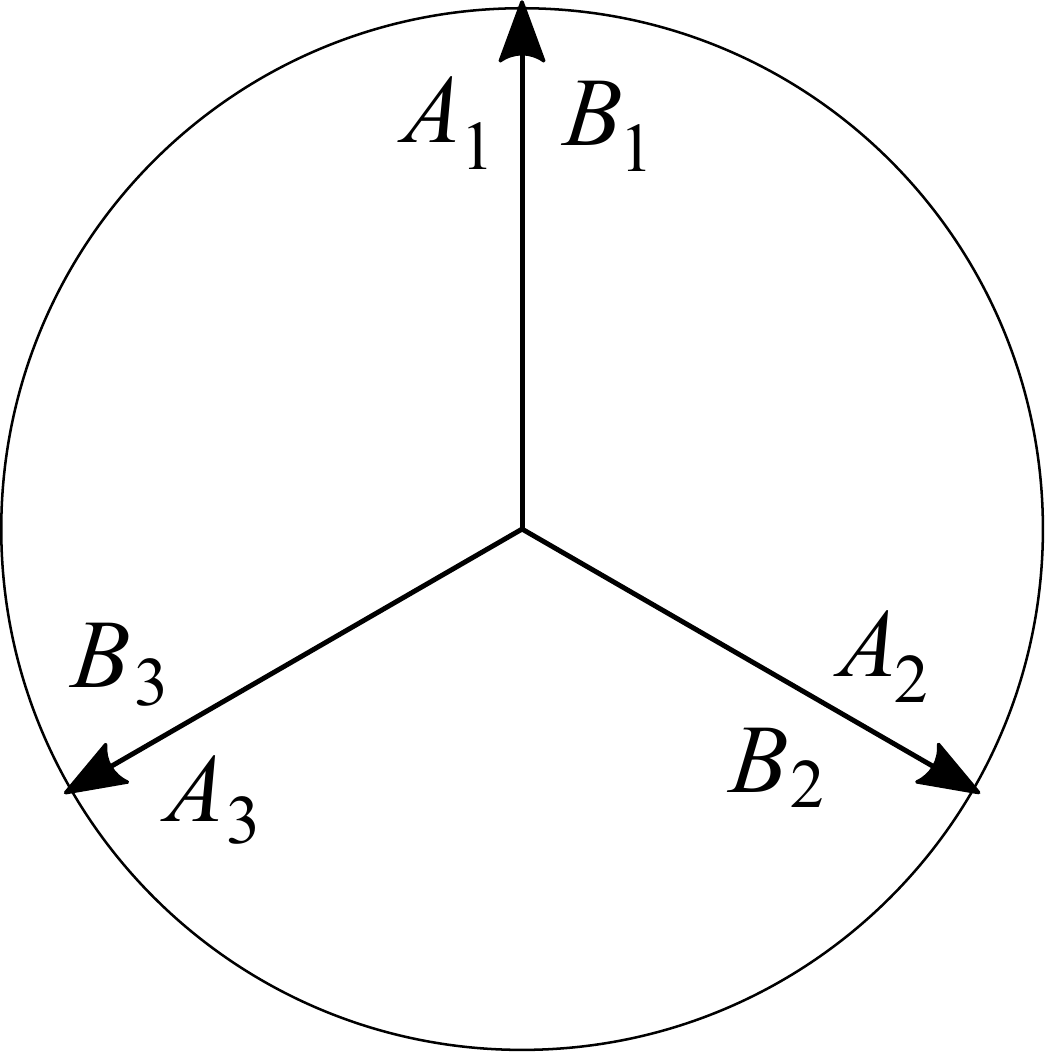}
  \caption{Distribution of projection axes $\hat{A}_x$ and $\hat{B}_y$ for the Bell state (\ref{bel}) on the equator of Bloch sphere -- the angle $\theta$ on the circle maps to the state $(e^{i\theta}|+\rangle+|-\rangle)/\sqrt{2}$.}
  \label{mom33}
\end{figure}

Let us consider the quantum case. The standard Bell state (maximally entangled)
\begin{equation}
\sqrt{2}|\psi\rangle=|+-\rangle-|-+\rangle \label{bel}
\end{equation}
and operators in $(|+\rangle,\;|-\rangle)$ bases
\begin{equation}
\hat{A}_x=\frac{1}{2}\begin{pmatrix}
1&e^{2\pi ix/3}\\
e^{-2\pi i x/3}&1\end{pmatrix}
\end{equation}
for $x=1,2,3$
(similarly $\hat{B}_y$) then
\begin{equation}
\langle A_xB_y\rangle=\langle A^2_xB^2_y\rangle=(1-\cos(2\pi(x-y)/3))/4
\end{equation}
The operators are in fact projections along regularly distributed axes on the equator of Bloch sphere, see Fig. \ref{mom33}.
In our case $\langle A_zB_z\rangle=0$ while $\langle A_xB_y\rangle=3/8$ for $x\neq y$ and the inequality is violated
with the left hand side equal $9/8$ while the right hand side is $2(9/8)-1=10/8>9/8$. The violation can be also quickly understood from the fact that 
$\langle A_z^2B_z^2\rangle=0$ implies that $A_zB_z=0$ so either $A_z=0$ or $B_z=0$ for each $z$, giving a simpler inequality
\begin{equation}
\langle A_1^2B_2^2\rangle+\langle A_2^2B_3^2\rangle+\langle A_3^2B_1^2\rangle+1\geq 2(\langle A_1B_2\rangle+\langle A_2B_3\rangle+\langle A_3B_1\rangle),
\end{equation}
checked by examining all cases, e.g. if $A_1=A_2=0$ then it reduces to $\langle A_3^2B_1^2\rangle+1\geq 2\langle A_3B_1\rangle$ obviously satisfied. Note that this inequality is only a restricted version of (\ref{in33}), valid if $A_zB_z=0$, $z=1,2,3$. Moreover, the fact that $\langle A^2_zB^2_z\rangle=0$ makes it clear that (\ref{in33}) does not satisfy property (ii) of (\ref{cla}).

In experimental practice, tests of local realism often cope with the null outcome, i.e. both observers register 0 or null -- a special outcome if no detection is registered -- at a low rate of production of entangled states. It happens e.g. in Clauser-Horne-Eberhard inequality \cite{chineq,eber}, which helps to take into account the finite efficiency of photon detectors. Note that the event
with only one observer registers null cannot be removed. Otherwise one has to assume fair sampling, which opens a loophole  for local realism.

Suppose the probability is dominated by the null event $A=B=0$ so that $\tilde{p}\to r\tilde{p}$
with $r$ being the (small) entanglement rate and $1-r$ being the probability of the null event.
Then the example (\ref{in33})
scales down everything except $-1$ on the right hand side at small entanglement production rate and violation disappears. We can get rid of  the null event by redefining $A'_1=1-A_1$, $B'_2=1-B_2$
when the inequality reads
\begin{align}
&\langle (1-A'_1)^2(1-B'_2)^2\rangle+\langle A_2^2B_3^2\rangle+\langle A_3^2B_1^2\rangle+\nonumber\\
&2\sqrt{\langle A_1^2B_3^2\rangle\langle A_2^2(1-B'_2)^2\rangle}+2\sqrt{\langle A_2^2B_1^2\rangle\langle A_3^2B_3^2\rangle}\nonumber\\
&+2\sqrt{\langle A_3^2B_2^2\rangle\langle (1-A'_1)^2B_1^2\rangle}\geq \nonumber\\
&2(\langle (1-A'_1)(1-B'_2)\rangle+\langle A_2B_3\rangle+\langle A_3B_1\rangle)-1,\label{in33r}
\end{align}
where the free terms (numbers) cancel at both sides. Thanks to the cancellation the inequality keeps being violated when non-null probability is scaled by $r$. Operationally the change of variables corresponds to taking complementary projection.

\section{Non-maximally entangled state -- two choices}

To get violation of a classical inequality with two choices for each observer, we will need a non-maximally entangled state.
The inequality reads in this case
\begin{align}
&\langle A_1^2B_2^2\rangle+\langle B_1^2A_2^2\rangle+\langle(A_1+B_1)^2\rangle/4
+\nonumber\\
&\sqrt{\langle(A_1-B_1)^2\rangle}\left(\sqrt{\langle A_1^2B_2^2\rangle}+\sqrt{\langle A_2^2B_1^2\rangle}\right)\nonumber\\
&+2\sqrt{\langle A_1^2B_1^2\rangle\langle A_2^2B_2^2\rangle}\geq 2(\langle A_1^2B_2\rangle+\langle B_1^2A_2\rangle).
\label{ine22}
\end{align}
The inequality does not belong to the class (\ref{cla}) because it does not satisfy its properties (i),(ii), and (iii).

We prove it starting from 
\begin{equation}
(A_1B_2+B_1A_2-(A_1+B_1)/2)^2\geq 0
\end{equation}
expanded into
\begin{align}
&\langle A_1^2B_2^2\rangle+\langle B_1^2A_2^2\rangle+2\langle A_1B_1A_2B_2\rangle+\langle(A_1+B_1)^2\rangle/4\nonumber\\
&+\langle(A_1-B_1)(A_1B_2-A_2B_1)\rangle
\geq 2(\langle A_1^2B_2\rangle+\langle B_1^2A_2\rangle).
\end{align}
Using CBS inequality
\begin{equation}
\sqrt{\langle A_1^2B_1^2\rangle\langle A_2^2B_2^2\rangle}\geq \langle A_1B_1A_2B_2\rangle
\end{equation}
and
\begin{eqnarray}
&&\sqrt{\langle(A_1-B_1)^2\rangle}\left(\sqrt{\langle A_1^2B_2^2\rangle}+\sqrt{\langle A_2^2B_1^2\rangle}\right)\nonumber\\
&&\geq\langle(A_1-B_1)(A_1B_2-A_2B_1)\rangle,
\end{eqnarray}
we get (\ref{ine22}).

\begin{figure}[t]
  \centering
  \includegraphics[width=.5\columnwidth]{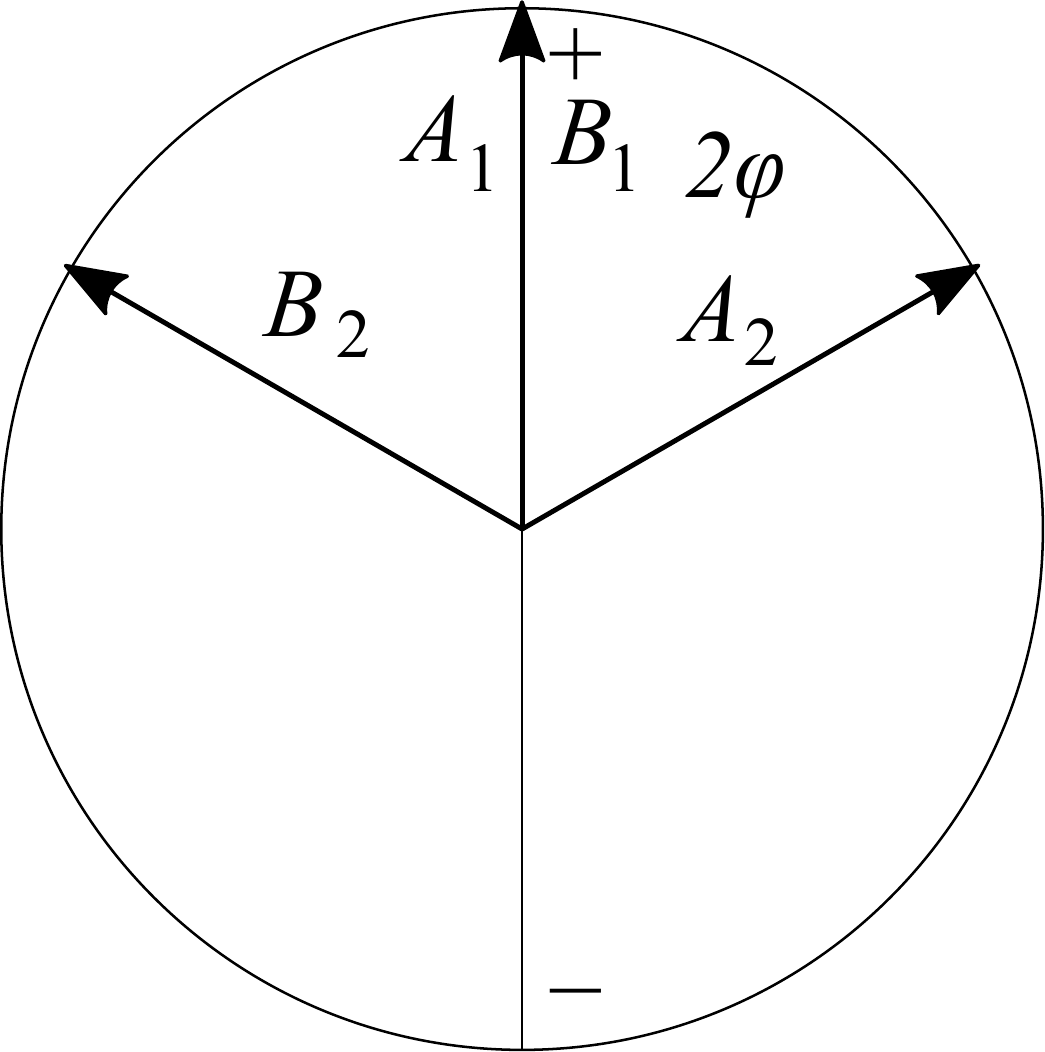}
  \caption{Distribution of projection axes $\hat{A}_x$ and $\hat{B}_y$ for the state (\ref{mm22}) on the opposite meridians of Bloch sphere ($\theta\to \cos(\theta/2)|+\rangle+\sin(\theta/2)|-\rangle$) to violate (\ref{ine22}).}
  \label{mom22}
\end{figure}
\begin{figure}[t]
  \centering
  \includegraphics[width=\columnwidth]{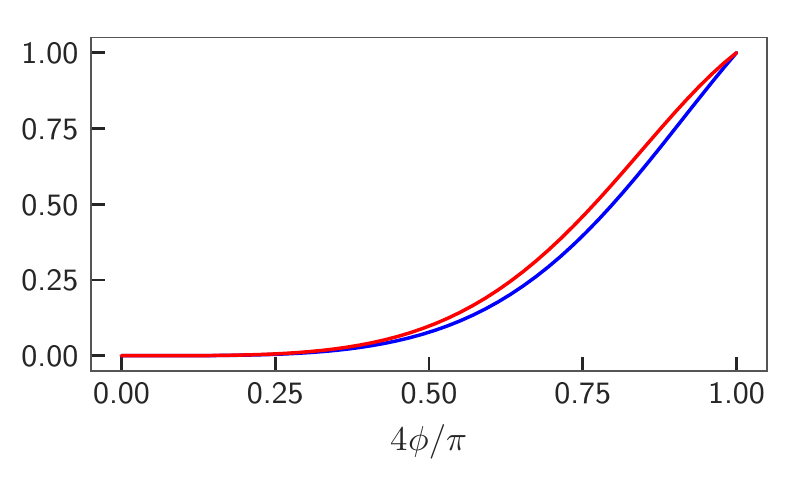}
  \caption{Violation of inequality (\ref{ine22}) for the state (\ref{mm22}) and operators depending on $\phi$ (see text), left hand side -- blue/lower, right hand side -- red/upper. Note that the curves differ only a little, and the difference disappears at $\phi=0$ (product state) or $\phi=\pi/4$ (maximally entangled state)}
  \label{vip}
\end{figure}

Let us take  $\hat{A}_1=\hat{B}_1=|+\rangle\langle +|$ and $\hat{A}_2=|n_+\rangle\langle n_+|$, $\hat{B}_2=|n_-\rangle\langle n_-|$
with $|n_\pm\rangle= \cos\phi|+\rangle\pm\sin\phi|-\rangle$, Fig. \ref{mom22}.
and the state 
\begin{eqnarray}
&&|\psi\rangle=\alpha|++\rangle+\beta|--\rangle,\label{mm22}\\
&&\alpha=\frac{\sin^2\phi}{\sqrt{\sin^4\phi+\cos^4\phi}},\:\beta=\frac{\cos^2\phi}{\sqrt{\sin^4\phi+\cos^4\phi}}.\nonumber
\end{eqnarray}
We have
\begin{align}
&\langle A_1^jB_1^k\rangle=\alpha^2\mbox{ for }j+k\geq 1,\nonumber\\
&\langle A_2^jB_2^k\rangle=0\mbox{ for }j,k\geq 1,\nonumber\\
&\langle A_2^j\rangle=\langle B_2^k\rangle=\alpha^2\cos^2\phi+\beta^2\sin^2\phi,\\
&\langle A_1^jB_2^k\rangle=\langle B_1^jA_2^k\rangle=\alpha^2\cos^2\phi\mbox{ for }j,k\geq 1.\nonumber
\end{align}
Then the inequality reads $\alpha^2(2\cos^2\phi+1)\geq 4\alpha^2\cos^2\phi$ which is violated whenever $\cos^2\phi>1/2$, i.e. $\phi<\pi/4$, although the violation is quite weak, see Fig. \ref{vip}.
Note also that the violation disappears when the the state becomes either maximally entangled or a simple product.

Again the violation is quickly understood from the fact that $\langle (A_1-B_1)^2\rangle=0$ together with $\langle A_1^2(1-B_1)^2\rangle=0$ implies $A_1=B_1=0,1$,
and $\langle A_2^2B_2^2\rangle$ implies $A_2=0$ or $B_2=0$.
In the case $A_1=B_1=1$, we have a simpler inequality $\langle B_2^2\rangle+\langle A_2^2\rangle +1\geq 2(\langle B_2\rangle+\langle A_2\rangle)$
which is true in both cases (either $A_2=0$ or $B_2=0$). Comparing with the previous section, the presented example is already robust against low entanglement rate 
(dominating null event) as all terms scale equally with non-null probability.

\section{Discussion and outlook}

We have shown that second-second moments suffice to reject local realism for two observers, with inequalities (\ref{in33}), (\ref{in33r}) and (\ref{ine22}). However, each observer has to use at least 3 choices for a maximally entangled state. Two choices suffice for a non-maximally entangled state but the proposed example is complicated while the violation is very weak. We suggest several further routes of research:
\begin{itemize}
\item Find an example with a larger violation.
\item Find violation by position and momentum or prove the impossibility.
\item Determine the class of inequalities which hold both in classical and quantum mechanics.
\item Apply these or new examples to realistic setup, adjusting if necessary.
\end{itemize}
Low-order moments can help to combine tests of local realism with relativity, which need a careful treatment of divergences in high-order correlations function. In the case of weak measurement, a larger violation should help to reduce the effect of background noise, which has to be subtracted from the statistics. Due to the very small violation in the presented examples, it is also important to check how much noise added to the outcome distribution spoils the violation in particular cases.

\section*{Acknowledgements}
I thank Witold Bednorz for mathematical consulting, Antonio Acin, and Wolfgang Belzig for discussions.

\appendix

\section{Positive polynomial not being a sum of polynomial squares}
\label{appa}

We will show that (\ref{nse}) is nonnegative. Changing variables
\begin{equation}
\sqrt{2}A_\pm=A_1\pm A_2,\: \sqrt{2}B_\pm=B_1\pm B_2
\end{equation}
the polynomial $W$ reads
\begin{eqnarray}
&&A_+^2+A_-^2+B_+^2+B_-^2+(A_+^2+A_-^2)(B_+^2+B_-^2)\nonumber\\
&&-3\sqrt{3/2}A_+B_+(A_-+B_-).
\end{eqnarray}
Denoting 
\begin{equation}
A=\sqrt{A_+^2+A_-^2}=\sqrt{A_1^2+A_2^2},\;B=\sqrt{B_+^2+B_-^2}=\sqrt{B_1^2+B_2^2},
\end{equation}
we have
\begin{equation}
W=(A^2+B^2)+A^2B^2-3\sqrt{3/2}A_+B_+(A_-+B_-).
\end{equation}
From H{\"o}lder inequality
\begin{eqnarray}
&&(A_-+B_-)^2=\left(A\frac{A_-}{A}+B\frac{B_-}{B}\right)^2\leq (A^2+B^2)\left(\frac{A_-^2}{A^2}+\frac{B_-^2}{B^2}\right)\nonumber\\
&&=(A^2+B^2)\left(2-\frac{A_+^2}{A^2}-\frac{B_+^2}{B^2}\right).
\end{eqnarray}
We have also
\begin{equation}
4(A_+B_+)^2=4A^2B^2\frac{A_+^2}{A^2}\frac{B_+^2}{B^2}\leq A^2B^2\left(\frac{A_+^2}{A^2}+\frac{B_+^2}{B^2}\right)^2
\end{equation}
so
\begin{eqnarray}
&&
(A_+B_+(A_-+B_-))^2\leq (A^2+B^2)A^2B^2t^2(2-t)/4\nonumber\\
&&\leq A^2B^2(A^2+B^2)8/27,
\end{eqnarray}
where $t=A_+^2/A^2+B_+^2/B^2\geq 0$ and we used the fact that the maximum of $t^2(2-t)$ for $t\geq 0$ is at $t=4/3$ and equal $32/27$.
Therefore,
\begin{equation}
|A_+B_+(A_-+B_-)|\leq (2/3)^{3/2}AB\sqrt{A^2+B^2}
\end{equation}
while
\begin{equation}
A^2+B^2+A^2B^2\geq 2\sqrt{A^2+B^2}AB
\end{equation}
completing the proof.

We will show that the polynomial cannot we written as $\sum_j Q_j^2$ where $Q_j(A_1,A_2,B_1,B_2)$ are polynomials. Equivalently
$Q_j$ can be polynomials of $A_\pm$, $B_\pm$ (change is linear)
but it can contain only $A_\pm$, $B_\pm$, $A_\pm B_\pm$, $A_\pm B_\mp$. Reducing quadratic form by standard methods we can arrange that only $Q_1$ contains $A_+$,
\begin{equation}
Q_1=A_+-\alpha A_-B_+-\beta A_-B_-.
\end{equation}
Note that $Q_1$ cannot contain $A_-$, $B_\pm$ or $A_+B_\pm$ because otherwise $Q_1^2$ would produce terms $A_+A_-$, $A_+B_\pm$, and $A_+^2B_\pm$, which cannot be cancelled later. 
Rearranging remaining quadratic terms, only $Q_2$ contains $A_-$ 
\begin{equation}
Q_2=A_--\gamma A_+B_++\beta A_+B_-.
\end{equation}
As above, it cannot contain $B_\pm$ or $A_-B_\pm$ while $-\beta$ term follows from the fact that $W$ does not contain $A_+A_-B_-$ which can appear only in $Q_1^2$ and $Q_2^2$.
Continuing rearranging, only $Q_3$ contains $B_+$ and only $Q_4$ contains $B_-$ so
\begin{equation}
Q_3=B_+-\delta B_-A_+-\eta B_-A_-,\;Q_4=B_--\xi B_+A_++\eta B_+A_-.
\end{equation}
Moreover $
\alpha+\gamma=(3/2)^{3/2}=\delta+\xi$
while
\begin{equation}
\sum_j Q_j^2=\alpha^2A_-^2B_+^2+\delta^2B_-^2A_+^2+(\gamma^2+\xi^2)A_+^2B_+^2+...,
\end{equation}
where the dotted term can only increase the first terms. On the other hand $W$ puts constraints
\begin{equation}
\alpha^2\leq 1,\: \delta^2\leq 1,\: \gamma^2+\xi^2\leq 1
\end{equation}
giving
$
\alpha^2+\gamma^2+\delta^2+\xi^2\leq 3
$
while
$
\alpha^2+\gamma^2\geq (\alpha+\gamma)^2/2=(3/2)^3/2
$
and the same for $\alpha\to \delta$, $\gamma\to \xi$. This would lead to $
(3/2)^3\leq 3
$
which is not true.

\section{Maximally entangled state and two choices}
\label{appb}

We will show that, counter-intuitively,  two choices $A_\pm$ are insufficient in the case of maximally entangled states,
i.e. there exists $\tilde{p}$ reproducing moments up to second-second order in agreement with quantum predictions. In Schmidt decomposition (\ref{schm}), a maximally entangled state is for  $\psi_j=1/\sqrt{N}$ with $j=1..N$

Both $\hat{A}_{+,-}$ and $\hat{B}$ (we postpone the generalization to many $B_y$ to the end of the proof) can have dimension larger than $N$. Let us us the block notation
\begin{equation}
\hat{B}\to \begin{pmatrix}\hat{B}_0&\hat{B}^\dag_e\\\hat{B}_e&\ast\end{pmatrix},
\end{equation}
with $\hat{B}_0$ restricted to the space of $1..N$.
Firstly, we make a diagonalization of $\hat{A}_\pm=\sum_{a_\pm}a_\pm|a_\pm\rangle\langle a_\pm|$.
We \emph{define} a joint probability (semipositive)
\begin{equation}
\tilde{p}(a_+,a_-)= |\langle a_+|\hat{1}_N|a_-\rangle|^2/N
\end{equation}
where $\hat{1}_N=\sum_j|j\rangle\langle j|$ i.e. it is projection to the space $1..N$. 
Our aim is to define positive conditional probability 
\begin{equation}
\tilde{p}(b|a_+,a_-)=\frac{\tilde{p}(b,a_+,a_-)}{\tilde{p}(a_+,a_-)}
\end{equation} 
for the cases $\tilde{p}(a_+,a_-)>0$ ($\tilde{p}(b,a_+,a_-)=0$ if $\tilde{p}(a_+,a_-)=0$) giving correct $\langle B\rangle_{a_\pm}$ and $\langle B^2\rangle_{a_\pm}$ defined as 
\begin{equation}
\langle B^k\rangle_{a_\pm}=\langle a_\pm|\hat{1}_N\hat{B}^{\ast k}\hat{1}_N|a_\pm\rangle/N=\sum_{b,a_\mp} b^kp(b,a_+,a_-).
\end{equation}
Here $\hat{B}$ is Hermitian and $\hat{B}^\ast=\hat{B}^T$ means either complex conjugation or transpose (equivalent).
If suffices to define moments $\langle b^k\rangle_{a_+,a_-}=\sum_bb^k p(b,a_+,a_-)$ for $k=1,2$
that satisfy
\begin{equation}
\langle b\rangle_{a_+,a_-}^2\leq \langle b^2\rangle_{a_+,a_-}p(a_+,a_-),\:\langle B^k\rangle_{a_\pm}=\sum_{a_\mp}\langle b^k\rangle_{a_+,a_-}\label{beineq}
\end{equation}
because then a positive Gaussian model
\begin{align}
&\tilde{p}(b|a_+,a_-)=\frac{p(a_+,a_-)}{\sqrt{2\pi(\langle b^2\rangle_{a+,a_-}\tilde{p}(a_+,a_-)-\langle b\rangle_{a_+,a_-}^2)}}\times\nonumber\\
&\exp\left(-\frac{(b\tilde{p}(a_+,a_-)-\langle b\rangle_{a_+,a_-})^2}{2\tilde{p}(a_+,a_-)(\langle b^2\rangle_{a+,a_-}\tilde{p}(a_+,a_-)-\langle b\rangle_{a_+,a_-}^2)}\right)
\end{align}
explains up to second-second moments.  The Gaussian distribution is only one of options, other choices include e.g. dichotomic distribution centered at the average. In the case of equality on (\ref{beineq}) 
we have $\tilde{p}(b|a_+,a_-)=\delta(b-\langle b\rangle_{a_+,a_-}/p(a_+,a_-))$.
We define 
\begin{eqnarray}
&&2N\langle b\rangle_{a_+,a_-}=\langle a_+|\hat{1}_N| a_-\rangle\langle a_-|\hat{1}_N\hat{B}^{\ast }\hat{1}_N|a_+\rangle
+\nonumber\\
&&\langle a_-|\hat{1}_N|a_+\rangle\langle a_+|\hat{1}_N\hat{B}^{\ast }\hat{1}_N|a_-\rangle,
\end{eqnarray}
which gives correct $\langle B\rangle_{a_\pm}$ by the fact that $\sum_{a_\mp}|a_\mp\rangle\langle a_\mp|$ is identity in the space containing $1..N$
(it does not matter if and how larger). 
We also define
\begin{equation}
\langle b^2\rangle_{0,a_+,a_-}=|\langle a_-|\hat{1}_N\hat{B}^{\ast }\hat{1}_N|a_+\rangle|^2/N
\end{equation}
which gives correct $\langle B_0^2\rangle_{a_\pm}$ analogously.
Moreover
\begin{equation}
\langle b\rangle_{a_+,a_-}^2\leq \langle b^2\rangle_{0,a_+,a_-}p(a_+,a_-)
\end{equation}
by the fact that 
\begin{eqnarray}
&&|\langle a_\pm|\hat{1}_N|a_\mp\rangle\langle a_\mp|\hat{1}_N\hat{B}^\ast\hat{1}_N|a_\pm\rangle|^2\leq\nonumber\\
&&
\langle a_\pm|\hat{1}_N \hat{B}^\ast\hat{1}_N|a_\pm\rangle\langle a_\mp|\hat{1}_N|a_\mp\rangle,
\end{eqnarray}
which follows from CBS inequality 
\begin{equation}
|\langle v|w\rangle\langle w|u\rangle|^2\leq \langle w|w\rangle^2\langle v|v\rangle\langle u|u\rangle
\end{equation}
(twice $|\langle s|t\rangle|^2\leq \langle s|s\rangle\langle t|t\rangle$ for $st=uw,wu$)
applied to
\begin{equation}
|v\rangle=\hat{1}_N|a_\pm\rangle,\:|w\rangle=\hat{1}_N|a_\mp\rangle,\:|u\rangle=\hat{B}^\ast\hat{1}_N|a_\pm\rangle
\end{equation}
and the fact the $\langle v|v\rangle\langle w|w\rangle\leq 1$ 
( both $|a_\mp\rangle$ are the normalized base vectors, while $\hat{1}_N$ projects them into a subspace).

The full second moments contain
$\hat{1}_N\hat{B}^{\ast 2}\hat{1}_N=\hat{B}_0^{\ast 2}+\hat{C}$ with $\hat{C}=\hat{B}_e^T\hat{B}_e^\ast$ being a semipositive operator.
Let us define $c(a_\pm)=\langle a_\pm|\hat{C}|a_\pm\rangle/N\geq 0$. Note that $c=\sum_{a_\pm}c(a_\pm)=\sum_j\langle j|\hat{C}|j\rangle/N$
does not depend on $\pm$. Finally
\begin{equation}
\langle b^2\rangle_{a_+,a_-}=\langle b^2\rangle_{0,a_+,a_-}+c(a_+)c(a_-)/c,
\end{equation}
assuming $c>0$. If $c=0$ then $\hat{C}=0$ and $\langle b^2\rangle_{a_+,a_-}=\langle b^2\rangle_{0,a_+,a_-}$.
One can easily check that it gives the correct full moments, keeping the desired inequality satisfied so the probability $\tilde{p}(b,a_+,a_-)\geq 0$ exists.
For many $\hat{B}_y$ we simply define
\begin{equation}
\tilde{p}(\{b\},a_+,a_-)=\tilde{p}(a_+,a_-)\prod_y \tilde{p}(b_y|a_+,a_-),
\end{equation}
which completes the proof.

\bibliographystyle{plain}

\end{document}